# Creating topological polar structure in a nonpolar matter

Adeel Y. Abid[1,6]†, Yuanwei Sun,[1,6]†, Xu Hou[3]†, Xiangli Zhong[2]†, Congbing Tan[4]*, Ruixue Zhu[1,6], Haoyun Chen[3], Ke Qu [5,6], Yuehui Li[1,6], Mei Wu[1,6], Jingmin Zhang[6], Jinbin Wang[2], Kaihui Liu[7,9], Xuedong Bai[12], Dapeng Yu[7,9,10], Jie Wang [3,8]*, Jiangyu Li[5,11]* and Peng Gao[1,6,7]*

[1]International Center for Quantum Materials, Peking University, Beijing 100871, China.

[2]School of Materials Science and Engineering, Xiangtan University, Hunan Xiangtan 411105, China.

[3]Department of Engineering Mechanics, School of Aeronautics and Astronautics, Zhejiang University, 310027, Hangzhou, China

[4]Department of Physics and Electronic Science, Hunan University of Science and Technology, 411201, Hunan Xiangtan, China

[5]Shenzhen Key Laboratory of Nanobiomechanics, Shenzhen Institutes of Advanced Technology, Chinese Academy of Sciences, Shenzhen 518055, Guangdong, China.

[6]Electron Microscopy Laboratory, School of Physics, Peking University, Beijing 100871, China.

[7]Collaborative Innovation Centre of Quantum Matter, Beijing 100871, China.

[8]Key Laboratory of Soft Machines and Smart Devices of Zhejiang Province, Zhejiang University, 310027, Hangzhou, China

[9]State Key Laboratory for Artificial Microstructure and Mesoscopic Physics, School of Physics, Peking University, Beijing 100871, China.

[10]Shenzhen Key Laboratory of Quantum Science and Engineering, Shenzhen 518055, China.

[11]Department of Materials Science and Engineering, Southern University of Science and Technology, Shenzhen 518055, Guangdong, China

[12]Beijing National Laboratory for Condensed Matter Physics and Institute of Physics, Chinese Academy of Sciences, Beijing 100190, China;

†These authors contributed equally to the work.

Email: p-gao@pku.edu.cn; lijy@sustech.edu.cn; cbtan@xtu.edu.cn; jw@zju.edu.cn.

Nontrivial topological structures offer rich playground in condensed matter physics including fluid dynamics[1,2], superconductivity[3], and ferromagnetism[4,5], and they promise alternative device configurations for post-Moore spintronics and electronics[6,7]. Indeed, magnetic skyrmions are actively pursued for high-density data storage[8], while polar vortices with exotic negative capacitance[9] may enable ultralow power consumption in microelectronics. Following extensive investigations on a variety of magnetic textures including vortices[5,10], domain walls[11] and skyrmions[12] in the past decades, studies on polar topologies have taken off in recent years, resulting in discoveries of closure domains[13], vortices[14], and skyrmions[15] in ferroelectric materials. Nevertheless, the atomic-scale creation of topological polar structures is largely confined in a single ferroelectric system, PbTiO$_3$ (PTO) with large polarization, casting doubt on the generality of polar topologies and limiting their potential applications. In this work, we successfully create previously unrealized atomic-scale polar antivortices in the nominally nonpolar SrTiO$_3$ (STO)[16], expanding the reaches of topological structures and completing an important missing link in polar topologies. The work shed considerable new insight into the formation of topological polar structures, and offers guidance in searching for new polar textures.

Both spins and dipoles prefer alignment, and often form uniform patterns that are topologically trivial. Nontrivial topologies such as vortices may arise[17,18], as schematically shown in the Supplementary Fig. 1a, often resulted from delicate energetic balance in confined structures and leading to exotic properties[9,10,19,20]. This is usually more difficult in polar systems wherein dielectric anisotropy is much stronger than magnetic one[17], and there is tremendous energy penalty when polarization rotates to form polar topologies. As a result the atomic-scale polar textures such as closure domains[13,21], vortices[14,18,22], and skyrmions[15] have only been observed in confined PTO layer within appropriately designed (PTO)$_n$/(STO)$_m$ superlattice (n-unit cell (u.c.) thick PTO and m-u.c. thick STO), and it is important to examine if such polar topologies are general in dielectric systems, or rather an exception in PTO. Equally important is if other topological structures often observed in magnetism, such as antivortices as schematically shown in Supplementary Fig. 1b, exist in dielectric materials or not.

Following the groundbreaking work of Kosterlitz and Thouless, it is now well known that vortex-antivortex pair as schematically shown in Fig. 1a may form during Kosterlitz-Thouless transition[23], which substantially reduces the energy penalty arising from individual vortex and antivortex. Such a vortex-antivortex pair has indeed been observed in superconducting[3] as well as ferromagnetic[24,25] systems, and it requires a pair of vortices with identical orientation. In $(PTO)_n/(STO)_m$ superlattice, however, the neighboring vortices in each PTO layer are observed to possess opposite orientations[14,26], making the topology in between trivial. We thus turn our search for antivortex to STO sandwiched between two layers of PTO instead. This may appear counterintuitive at the first sight, though polar order has indeed been observed in STO before at the reduced thickness[16] or in a confined heterostrucure[27,28], and the weaker polarization induced in nominally cubic STO may exhibit weaker anisotropy, facilitating the formation of vortex-antivortex pair. Motivated by such considerations and guided by detailed phase field simulations, we design a series of $(PTO)_n/(STO)_m$ heterostructures, and successfully create vortex-antivortex pairs in $(PTO)_{10}/(STO)_4$ system. To our best knowledge, this is not only the first observation of atomic-scale polar antivortex, but also the first realization of polar topology in a nonpolar system, wherein topological phase transition can be induced by either temperature change or electric field with tunable dielectric properties. By examining the energetics of the superlattice, we conclude that the driving force for such antivortex formation is electrostatic, while misfit strain plays a negligible role, offering further insight into searching for new topological polar structures.

We first seek to create an antivortex as schematically shown in Fig. 1a, sandwiched between two vortices. Such topological structure has been predicted by Mermin from energetic point of view[29], though its realization has yet to be demonstrated in a dielectric system. We thus consider superlattices with configuration of $(PTO)_n/(STO)_m$, wherein m-u.c. thick STO layer is sandwiched between two n-u.c. thick PTO layers. Array of polar vortices has recently been observed in PTO layer of such superlattices[14], giving us hope that under appropriate design antivortex may emerge in STO sandwiched between two vortices in two adjacent PTO layers. Based on systematic phase field simulations, we have identified four typical polar configurations (Fig. 1b-e) for $(PTO)_n/(STO)_m$, enabling us to construct a phase diagram to guide the design (Fig. 1f). When both PTO and STO layers are ultrathin, for example for m, n=4, antiparallel $a$-domain is observed in two adjacent PTO layers, while polarization in STO layer is negligibly small, exhibiting no nontrivial topology (Fig. 1b). When they are both relatively thick, for example n=10 and m=20, nontrivial vortex array emerges in PTO, while polarization in STO remains negligibly small (Fig. 1c). We thus keep n=10 to maintain the desired vortex array in PTO, and reduce the thickness of STO. At m=10, sign of antivortex pattern appears in STO (Fig. 1d), with its polarization magnitude modestly increases, though the topological structure is not regular, and two vortices in adjacent PTO is not well aligned. When thickness of STO is further reduced to 4-u.c., regular antivortex emerges in STO (Fig. 1e), sandwiched between two nicely aligned vortices in PTO, and the magnitude of its polarization increases further as well. This is precisely what we are looking for, fully consistent with theoretical expectation illustrated in Fig. 1a. The window for the vortex-antivortex pair is quite narrow in the phase diagram (Fig. 1f), with thickness of PTO ranging between 8 to 12-u.c. and thickness of STO smaller than 8-u.c. Note that similar superlattices have been studied by Hong et al. [28], though their focus was polar configuration of PTO.

Encouraged by phase field simulation, we design a gradient superlattice heterostructure of $(PTO)_{10}/(STO)_m$, with thickness of PTO fixed at 10-u.c., while that of STO varying at 4, 7, 10, 15, as shown in Supplementary Fig. 2a. The superlattice heterostructures were then grown on $DyScO_3$ (110) substrate by pulsed laser deposition (PLD). The low-magnification high angle annular dark field (HAADF) scanning transmission electron microscopy (STEM) image of $(PTO)_{10}/(STO)_m$ heterostructure in Fig. 2a illustrates the stacked ferroelectric and dielectric layers of different contrast, whereas Supplementary Fig. 2b-d show distributions of strain components estimated from geometric phase analysis (GPA) based on the STEM image. The white-colored sinusoidal wave-like out-of-plane strain pattern is observed within PTO layers along the [100] direction, suggesting the existence of long-range vortex ordering consistent with previous reports[21,22,30]. The dark field transmission electron microscopy (TEM) image shown in Fig. 2b depicts periodic array of bright and dark intensity modulation, corresponding to the clockwise-anticlockwise vortex pairs previously reported in PTO layers[14,18,30]. We can also see such vortex ordering from the spatial distribution of polarization calculated from phase field simulation (Fig. 2c) that closely resembles Fig. 2b, wherein zoomed-in examination at the interface between dark and bright contrasts clearly reveals a polar vortex.

In order to confirm the polar structure in the superlattice at the atomic scale, we acquired high magnification HAADF image for 4-u.c. STO sandwiched between 10-u.c. PTO, as shown in Fig. 2d. The Z-contrast sensitivity of HAADF imaging shows sharp and coherent interfaces between PTO and STO (Z is the atomic number), which is also confirmed by the atomically resolved energy dispersive X-ray spectra (EDS) mapping incorporated in Supplementary Fig. 3. The polar map (Fig. 2e) of displacement vectors between A-site (Pb, Sr) and B-site (Ti) derived from HAADF image[31] illustrates a pair of antivortices within the STO layer, as highlighted by the dotted diamond boxes at their cores, and each antivortex is sandwiched between a pair of vortices in adjacent PTO layers, fully consistent with the theoretical expectation in Fig. 1a. To better appreciate the topology of polar structures, enlarged views of polarization vectors overlaid with polar angle variation for the marked rectangular boxes in PTO (Fig. 2f) and STO (Fig. 2g) are examined, revealing clearly vortex structure in PTO and antivortex in STO. Moreover, the variations of polar displacement within the antivortex along A-B (Fig. 2h) and C-D (Fig. 2i) directions show that out-of-plane ($D_z$) and in-plane ($D_x$) polar vectors reverse their directions when passing through the antivortex core, approaching and departing the core from two sets of opposite directions (head-to-head and tail-to-tail)[25,27,29,32]. Using the experimental data, we also obtain the distribution of winding numbers[33] (Supplementary Fig. 4a), confirming the topological nature of vortices and antivortices about their respective cores, where 2 antivortices with winding number -1 between 4 vortices with winding number 1 are revealed. Additional details on the polar topologies can be found in Supplementary Fig. 4b-d along with phase field simulations (Supplementary Fig. 4e-g), which show good agreement between experiment and simulation.

The accurate quantitative measurement of polarization in STO, particularly at larger thicknesses, remains a challenge for HAADF image because the polarity in STO mainly arises from the displacement of the oxygen[16], while HAADF tends to underestimate the STO polarization relative to that of PTO[34]. Thus we also acquired integrated differential phase contrast (iDPC) image[22], which presents the information of oxygen configurations with picometer precision and thus gives better accuracy for polarization measurements based on the atomic displacements between cations and oxygen[22] (see Methods for details). From the iDPC image (colored for clarity) in Fig. 3a, the atomic shift between Sr and O for 4-u.c. thick STO with respect to their respective centrosymmetric positions is up to ~ 20 pm (Supplementary Fig. 5,6), visible even with the naked eye. From the enlarged views of atomic structure shown in the inset of Fig. 3a, the octahedron shift in STO (red) are similar to that of PTO (shallow yellow) except less pronounced. The corresponding polar map in Fig. 3b illustrates three antivortices in the STO layer, with their cores highlighted by the dotted boxes. When passing through one of the antivortex cores along either A-B or C-D direction as marked, the polar vectors reverse their directions (Fig. 3c,d) in a similar manner as already revealed by HAADF image (Fig. 2h,i), demonstrating high fidelity of our analysis based on two independent techniques and data sets. The atomic structure of the polarized STO exhibits larger displacement between cations and O and smaller one between Sr and Ti, similar to that of PTO[35], though its magnitudes of displacements and thus polarization are much smaller (see Supplementary Fig. 6). Indeed, the polarization in the polarized STO is mainly contributed by the oxygen displacement, in good agreement with the previous study[16]. Here the average magnitude of the polarization is estimated to be ~30 $\mu C/cm^2$ (Supplementary Fig. 6g), consistent with phase field simulation as well as previous first principle density functional theory (DFT) calculations[27]. With increased STO thickness, the polarization decreases (Fig. 3e and Supplementary Fig. 7), with the antivortex-like polar topology remaining for 7-u.c. STO, though the structure is less ideal (Supplementary Fig. 8).

These two sets of indepent STEM data acquired using HAADF and iDPC techniques unambiguously established the existence of antivortex topology in STO, and it is compelling for us to examine its energetics, as shown in Fig. 4a, so that we can understand its formation mechanism. Interestingly, the electrostatic energy density in STO is found from phase field simulation to be negative and decrease with reduced STO thickness, while the corresponding elastic energy is positive and does not change much with STO thickness. It suggests that the formation of antivortex in STO is largely driven by electric field, while misfit strain in superlattice plays a negligible role. This is in sharp contrast to corresponding analysis for PTO (Supplementary Fig. 9) showing that elastic energy is negative while electric energy is positive, so that the driving force for vortex formation in PTO is elastic, as commonly understood. An immediate implication of this finding is that we may be able to tune the antivortex electrically[36], as exhibited by the hysteresis loop of winding number versus electric field in Fig. 4b. Reversible field induced topological phase transition is observed, where the vortex-antivortex pair is turned into a single domain state upon a modest electric field around 392 kV/cm, and is recovered when the electric field is reduced to 209 kV/cm. Similar phase transition can also be induced by heating and cooling (Supplementary Fig. 10), analogue to Kosterlitz-Thouless transition[23]. The electric tuning also enables us to examine the stability of antivortex in STO with different thickness, measured by the critical electric field for topological phase transition (Fig. 4a). As expected, this critical field is largest for 4-u.c. STO, demonstrating its highest stability, while that of 10-u.c. STO is substantially reduced and thus is much less stable. Furthermore, the antivortex exhibits positive and much enhanced capacitance at its core, as shown in Fig. 4c, while vortex possesses negative capacitance in excellent agreement with previous report[9] (Supplementary Fig. 11). Interestingly, the field induced phase transition renders dielectric hysteresis as shown in Fig. 4d, where dielectric tunability as large as 50.7% is observed.

The vortex-antivortex pair we create is best illustrated in Supplementary Fig. 12, which highlights the importance of electric boundary conditions for delicate antivortex energetics. Such a delicate balance together with small and highly nonuniform polarization may explain why all previous studies did not notice its existence in STO. Our work completes an important missing link in polar topology, where stable antivortex is finally confirmed to exist at atomic scale after the discoveries of polar vortex and skyrmions. More importantly, it expands the polar topologies beyond not only PTO, but also beyond ferroelectrics in general, pointing toward a direction for designing new polar topologies in dielectric systems. The energetics of such vortex-antivortex pair is rather delicate, making it nontrivial to realize experimentally, but also make it easier to manipulate via external mechanisms that can easily tip the energetic balances, as we have demonstrated. Our study thus offer a realistic roadmap forward to ultimately engineer and control the polar topologies for devices applications.

**Methods**

**Fabrication of designed gradient heterostructures.** Superlattices of $(PTO)_n/(STO)_m$ were deposited on $SrRuO_3$ (SRO)-buffered, (110)-$DyScO_3$ substrates in a PLD system (PVD-5000) equipped with a KrF excimer laser ($\lambda$=248 nm). Ceramic targets of SRO, STO and $Pb_{1.1}TiO_3$ (10 mol% excessive amount of lead to compensate the evaporation loss of Pb) were used for the PLD deposition of bottom electrode and the $(PTO)_n/(STO)_m$ superlattices, respectively. The SRO and superlattices were grown under a laser energy of 390 and 340 mJ pulse$^{-1}$, respectively, with a pulse repetition rate of 10 Hz. The SRO was firstly deposited at 690 °C in an 80 mtorr oxygen pressure, and then the substrate was cooled down to 600 °C for the subsequent growth of the $(PTO)_n/(STO)_m$ superlattices. The growth temperature and oxygen pressure for the growth of the superlattices was 600 °C and 200 mtorr, respectively. By controlling the growth time, thicknesses of the SRO, PTO, and STO layers were held at desired thickness. Right after growth of the $(PTO)_n/(STO)_m$ superlattices, the samples were cooled down to room temperature in 200 mtorr oxygen pressure.

**TEM cross-sectional sample preparation.** For image acquisition the cross-sectional TEM specimen were thinned to less than 30 µm first by using mechanical polishing. The subsequent argon ion milling was carried out using PIPS$^{TM}$ (Model 691, Gatan Inc.) with the accelerating voltage of 3.5 kV until a hole was made. Low voltage milling was performed with accelerating voltage of 0.3 kV to minimize damage and remove the surface amorphous layer.

**Electron microscopy characterization and image analysis.** The dark field transmission electron microscopy image shown in Fig. 2b was carried out under the two-beam condition with **g**-vector: $\mathbf{g} = 002_{pc}$ from an aberration-corrected FEI Titan Themis G2 at 300 kV. HAADF and iDPC images were also recorded at 300 kV using an aberration-corrected FEI Titan Themis G2. The convergence semi-angle for imaging is 30 mrad, the collection semi-angles snap is 4 to 21 mrad for the iDPC imaging and 39 to 200 mrad for the HAADF. The atom positions were determined by simultaneously fitting with two-dimensional Gaussian peaks using a MatLab code[31]. The polar vectors in Fig. 2e were plotted from the offset between A (Pb and Sr) and B-site (Ti) sublattices based on the HAADF-STEM image in Fig. 2d. To determine the atomic shift for each atom column in the HAADF, the displacements of A (B) with respect to the center of their surrounding four B (A) columns are measured and decomposed into in-plane and out-of-plane components, respectively. For the iDPC image along [010] direction (Fig. 3a), each cation column [Pb(Sr) and TiO] is surrounded by four oxygen columns. The displacements of cations with respect to the center of their surrounding four oxygen columns can be measured along in-plane and out-of-plane directions, receptively (Supplementary Figs. 5 and 6). Based on the displacements, the unit-scale polarization (Fig. 3b, e and Supplementary Fig. 6g) can be calculated according to $P_s = \frac{1}{V}\sum \delta_i Z_i$ [37]: where V is the volume of unit cell which for our case is $a^2c$, $\delta$ is displacement/shift of atom (*i*) from their centrosymmetric position and Z is the Born effective charge of atom (*i*) calculated by ab initio theory, 6.71 for Ti, 3.92 for Pb, -2.56 for O in PTO, and 7.12 for Ti, 2.54 for Sr, -2.00 for O in STO [37]. Taking the oxygen sublattice as the reference (standard position of each unit cell), the polarization is simplified to be $P_s = \frac{1}{V}(\delta_{Pb(Sr)-O}Z_{Pb(Sr)-O} + \delta_{Ti-O}Z_{Ti-O})$. The vector and magnitude maps of displacement (Fig. 2e) and polarization (Fig. 3b) are plotted by Origin.

**Winding number calculation**. We carried out a local winding number analysis for polar angle distribution in order to authenticate the topological nature of vortices and antivortices about their respective cores. The two-dimension winding number *n* along a closed loop *C* was calculated by the following line integral $n = \frac{1}{2\pi}\oint_C \nabla\theta \cdot \mathrm{d}r$, where $\nabla\theta$ is the angle gradient of polarization vectors along the integral loop $C$ [29,33,38]. By performing a local winding number calculation on each closed loop, we are able to identify the existence of a single vortex and antivortex that gives a winding number equal to +1 and -1 respectively (For illustration simultaneously observe Fig. 2 e, f, g, Fig. 4b, Supplementary Fig. 4 and Supplementary Fig. 10. We used a condition of $|\Delta\theta|<180°$ to determine the angle rotation direction. We have accomplished the winding number quantification for each loop by taking 2×2 square loops tilted on the vector map sharing the boundaries.

**Phase field simulation details:** In the phase field modelling, the spatially-dependent polarization vector ***P*** is selected as the order parameter to describe the polar states, and the total free-energy density of a PTO/STO superlattice thin film takes the following form:

$$f = \alpha_i P_i^2 + \alpha_{ij} P_i^2 P_j^2 + \alpha_{ijk} P_i^2 P_j^2 P_k^2 + \frac{1}{2} c_{ijkl}\varepsilon_{ij}\varepsilon_{kl} - q_{ijkl}\varepsilon_{ij}P_k P_l$$
$$+ \frac{1}{2} g_{ijkl}(\partial P_i/\partial x_j)(\partial P_k/\partial x_l) - \frac{1}{2}\varepsilon_0\varepsilon_r E_i E_i - E_i P_i, \tag{1}$$

where $\alpha_i$, $\alpha_{ij}$ and $\alpha_{ijk}$ are the Landau expansion coefficients (sixth- and fourth-order forms for PTO and STO, respectively), $c_{ijkl}$ is the elastic constant, $q_{ijkl}$ is the electrostrictive coefficient, $g_{ijkl}$ is the gradient energy coefficient, $\varepsilon_0$ is the dielectric constant of vacuum, and $\varepsilon_r$ denotes the relative dielectric constant of the background material (cubic PTO and STO in this case). The summation convention for the repeated indices is employed, and the Latin letters $i, j, k, l$ take $1, 2$ in the present work. The detailed expression of each Landau energy forms can be found in the literature[39]. Based on the total free-energy density, the temporal evolution of the polarization field can be obtained by solving the time-dependent Ginzburg-Landau (TDGL) equation:

$$\frac{\partial P_i(\boldsymbol{r}, t)}{\partial t} = -L \frac{\delta F}{\delta P_i(\boldsymbol{r}, t)} (i = 1, 2), \tag{2}$$

where $L$ represents the domain wall mobility, $F = \int_V f \, dV$ is the total free-energy, $\boldsymbol{r}$ is the spatial position vector, and $t$ denotes time. Besides the TDGL equation, both the mechanical equilibrium equation,

$$\sigma_{ij,j} = \partial(\partial f / \partial \varepsilon_{ij}) / \partial x_j = 0, \tag{3}$$

and the electrostatic equilibrium equation,

$$D_{i,i} = \partial(-\partial f / \partial E_i) / \partial x_i = q, \tag{4}$$

must be satisfied simultaneously for a ferroelectric system, where $q$ is the space charge.

To solve the above equations, the nonlinear finite element method and backward Euler iteration method are employed for space discretization and time integration, respectively. For clearer illustration and computational simplicity, the carried-out simulations were restricted to the [100]-[001] crystallographic plane, which corresponds to the x-z plane in the Cartesian coordinate system. Discrete grids with $\Delta x = \Delta z = 0.4$ nm in real space were used for space discretization, and the step length for time integration was chosen as $\Delta t / t_0 = 0.2$, where $t_0 = 1/(\alpha_0 L)$ and $\alpha_0$ is the absolute value of $\alpha_1$ at room temperature. Periodic boundary conditions for the electric potential and polarization components were employed along the x direction. The material parameters for PTO and STO used in the simulations can be found in the literature[39,40]. In addition, the isotropic gradient energy coefficient for both PTO and STO was chosen as $g_{11}/G_{110} = 0.4$, where $G_{110} = 1.73 \times 10^{-10} \text{ C}^{-2} \text{m}^4 \text{N}$. A normalization process for the material parameters was used to achieve better numerical stability in the simulations, which can be referred to elsewhere[41,42]. Small random fluctuation ($<0.01 P_0$, where $P_0 = 0.757$ C m$^{-2}$ is the spontaneous polarization of PTO at room temperature) was used as the initial setup for polarization to initiate its evolution. The cubic lattice constants for paraelectric PTO and STO were assumed as 3.955 Å and 3.905 Å, respectively, and the resulted interlayer mechanical inhomogeneity was taken into account in the modeling by exerting a misfit strain of -0.3% on the PTO layers.

The two-dimensional phase-field simulations were carried out for $(PTO)_n/(STO)_m$ superlattices at room temperature. First, corresponding to the configuration of the gradient $(PTO)_{10}/(STO)_m$ superlattice in the experiment, the phase-field simulations were used to calculate its polarization and strain distributions. In addition, the phase-field calculations were further conducted to complete a phase diagram for the evolution of antivortex states in $(PTO)_n/(STO)_m$ superlattices. The average polarization (blue color data in Fig. 3e) variation within the STO layer for $(PTO)_{10}/(STO)_m$ superlattice is calculated by $P_{avg} = (\sum_{i=0}^{N} P_i)/N$, where $P_i$ is the polarization magnitude of each node inside STO layer and $N$ is the total number of the nodes. For energy density calculation we utilized the formula $f_{avg} = (\sum_{i=0}^{M} f_i)/M$, where $f_i$ is the total energy density of each element inside $(PTO)_{10}/(STO)_m/(PTO)_{10}$ superlattice system wherein the subscripts m= 4, 6, 8 and 10-u.c as can be seen in Fig. 4a. Furthermore, $f_i$ is calculated by Gaussian integration based on the equation (1) and $M$ is total number of elements in the investigated $(PTO)_{10}/(STO)_m/(PTO)_{10}$ system. The local permittivity given in Fig. 4c and Supplementary Fig. 11 is calculated by $\varepsilon = \Delta E_z/\Delta P_z$ after applying a small electric field to the initial stable domain structure without electric field.

**References**

1       Kleckner, D. & Irvine, W. T. M. Creation and dynamics of knotted vortices. *Nat. Phys.* **9**, 253-258 (2013).

2       Donati, S. *et al.* Twist of generalized skyrmions and spin vortices in a polariton superfluid. *Natl. Acad. Sci. U. S. A.* **113**, 14926-14931 (2016).

3       Chibotaru, L. F., Ceulemans, A., Bruyndoncx, V. & Moshchalkov, V. V. Symmetry-induced formation of antivortices in mesoscopic superconductors. *Nature* **408**, 833-835 (2000).

4       Wang, L. *et al.* Ferroelectrically tunable magnetic skyrmions in ultrathin oxide heterostructures. *Nat. Mater.* **17**, 1087-1094 (2018).

5       Huang, F.-T. & Cheong, S.-W. Aperiodic topological order in the domain configurations of functional materials. *Nat. Rev. Mater.* **2**, 17004 (2017).

6       Bohr, M. T. & Young, I. A. CMOS scaling trends and beyond. *IEEE Micro* **37**, 20-29 (2017).

7       Sharma, P. *et al.* Nonvolatile ferroelectric domain wall memory. *Sci. Adv.* **3**, e1700512 (2017).

8       Nagaosa, N. & Tokura, Y. Topological properties and dynamics of magnetic skyrmions. *Nat. Nanotechnol.* **8**, 899-911 (2013).

9       Yadav, A. K. *et al.* Spatially resolved steady-state negative capacitance. *Nature* **565**, 468-471 (2019).

10      Wachowiak, A. *et al.* Direct Observation of Internal Spin Structure of Magnetic Vortex Cores. *Science* **298**, 577-580 (2002).

11      Parkin, S. S. P., Hayashi, M. & Thomas, L. Magnetic Domain-Wall Racetrack Memory. *Science* **320**, 190-194 (2008).

12      Yu, X. Z. *et al.* Real-space observation of a two-dimensional skyrmion crystal. *Nature* **465**, 901-904 (2010).

13      Tang, Y. *et al.* Observation of a periodic array of flux-closure quadrants in strained ferroelectric PbTiO$_3$ films. *Science* **348**, 547-551 (2015).

14      Yadav, A. K. *et al.* Observation of polar vortices in oxide superlattices. *Nature* **530**, 198-201 (2016).

15      Das, S. *et al.* Observation of room-temperature polar skyrmions. *Nature* **568**, 368-372 (2019).

16      Lee, D. *et al.* Emergence of room-temperature ferroelectricity at reduced dimensions. *Science* **349**, 1314-1317 (2015).

17      Tian, G. *et al.* Topological domain states and magnetoelectric properties in multiferroic nanostructures. *Natl. Sci. Rev.* **6**, 684-702 (2019).

18      Hsu, S. L. *et al.* Emergence of the Vortex State in Confined Ferroelectric Heterostructures. *Adv. Mater.* **31**, e1901014 (2019).

19      Balke, N. *et al.* Enhanced electric conductivity at ferroelectric vortex cores in BiFeO$_3$. *Nat. Phys.* **8**, 81-88 (2011).

20      Choe, S.-B. *et al.* Vortex Core-Driven Magnetization Dynamics. *Science* **304**, 420-422 (2004).

21      Liu, Y. *et al.* Large Scale Two-Dimensional Flux-Closure Domain Arrays in Oxide Multilayers and Their Controlled Growth. *Nano Lett.* **17**, 7258-7266 (2017).

22    Sun, Y. *et al.* Subunit cell–level measurement of polarization in an individual polar vortex. *Sci. Adv.* **5**, eaav4355 (2019).

23    Kosterlitz, J. M. & Thouless, D. J. Ordering, metastability and phase transitions in two-dimensional systems. *J. Phys. C: Solid State Phys.* **6**, 1181 (1973).

24    Kuepper, K., Buess, M., Raabe, J., Quitmann, C. & Fassbender, J. Dynamic vortex-antivortex interaction in a single cross-tie wall. *Phys. Rev. Lett.* **99**, 167202 (2007).

25    Ruotolo, A. *et al.* Phase-locking of magnetic vortices mediated by antivortices. *Nat. Nanotechnol.* **4**, 528-532 (2009).

26    Shafer, P. *et al.* Emergent chirality in the electric polarization texture of titanate superlattices. *Proc. Natl. Acad. Sci. U. S. A.* **115**, 915-920 (2018).

27    Aguado-Puente, P. & Junquera, J. Structural and energetic properties of domains in PbTiO$_3$/SrTiO$_3$ superlattices from first principles. *Phys. Rev. B* **85**, 184105 (2012).

28    Hong, Z. *et al.* Stability of Polar Vortex Lattice in Ferroelectric Superlattices. *Nano Lett* **17**, 2246-2252 (2017).

29    Mermin, N. D. The topological theory of defects in ordered media. *Rev. Mod. Phys.* **51**, 591-648 (1979).

30    Du, K. *et al.* Manipulating topological transformations of polar structures through real-time observation of the dynamic polarization evolution. *Nat. Commun.* **10**, 4864 (2019).

31    Nelson, C. T. *et al.* Spontaneous Vortex Nanodomain Arrays at Ferroelectric Heterointerfaces. *Nano Lett.* **11**, 828-834 (2011).

32    Tian, X., He, X. & Lu, J. Atomic scale study of the anti-vortex domain structure in polycrystalline ferroelectric. *Philos. Mag.* **98**, 118-138 (2017).

33    Kim, J., You, M., Kim, K.-E., Chu, K. & Yang, C.-H. Artificial creation and separation of a single vortex–antivortex pair in a ferroelectric flatland. *npj Quantum Mater.* **4**, 29 (2019).

34    Li, Q. *et al.* Quantification of flexoelectricity in PbTiO$_3$/SrTiO$_3$ superlattice polar vortices using machine learning and phase-field modeling. *Nat. Commun.* **8**, 1468 (2017).

35    Shirane, G., Pepinsky, R. & Frazer, B. C. X-Ray and Neutron Diffraction Study of Ferroelectric PbTiO$_3$. *Acta. Cryst.* **9**, 131 (1956).

36    Kim, K. E. *et al.* Configurable topological textures in strain graded ferroelectric nanoplates. *Nat. Commun.* **9**, 403 (2018).

37    Zhong, W., King-Smith, R. D. & Vanderbilt, D. Giant LO-TO splittings in perovskite

ferroelectrics. *Phys. Rev. Lett.* **72**, 3618-3621 (1994).

38      Trebin, H. R. The topology of non-uniform media in condensed matter physics. *Adv. Phys.* **31**, 195-254 (2006).

39      Chen, L. Q. APPENDIX A-Landau free-energy coefficients, Physics of Ferroelectrics. Springer, Berlin, Heidelberg. 363-372 (2007).

40      Li, Y. L. *et al*. Effect of substrate constraint on the stability and evolution of ferroelectric domain structures in thin films. *Acta. Mater.* **50,** 395-411 (2002).

41      Wang, J. *et al*. Phase-field simulations of ferroelectric/ferroelastic polarization switching. *Acta. Mater.* **52,** 749-764 (2004).

42      Wang, J. *et al*. Role of grain orientation distribution in the ferroelectric and ferroelastic domain switching of ferroelectric polycrystals. *Acta. Mater.* **61,** 6037-6049 (2013).

**Acknowledgements**

This research was supported by the National Key R&D Program of China (grant no. 2016YFA0300804, 2016YFA0201001, 2016YFA0300903), the National Natural Science Foundation of China (grant nos. 51672007, 11974023, 11875229, 51872251, 11972320 and 11672264), the National Equipment Program of China (ZDYZ2015-1), the Key R&D Program of Guangdong Province (grant nos. 2018B030327001, 2018B010109009, and 2019B010931001), Shenzhen Science and Technology Innovation Committee (JCYJ20170818163902553), Zhejiang Provincial Natural Science Foundation (Grant No. LZ17A020001) and the "2011 Program" Peking-Tsinghua-IOP Collaborative Innovation Center for Quantum Matter. We also acknowledge the Electron Microscopy Laboratory in Peking University for the use of the Cs-corrected electron microscope.

**Author Contributions**

PG, CBT, JW, and JYL conceived the idea and designed the work. CBT grew the samples assisted by JBW and XLZ. YWS performed the electron microscopy experiments and assisted by AYA, RXZ, KQ, and JMZ under the direction of PG. XH and HYC did the phase field simulation under the direction of JW and JYL. YS and AYA performed data analysis assisted by RXZ, MW and YHL under the direction of JW, JYL and PG. AYA first found the sign of antivortex in STEM image. JYL and PG wrote the manuscript with the assistance of AYA and YWS and all the authors. All authors discussed the results and commented on the manuscript.

**Figures and captions**

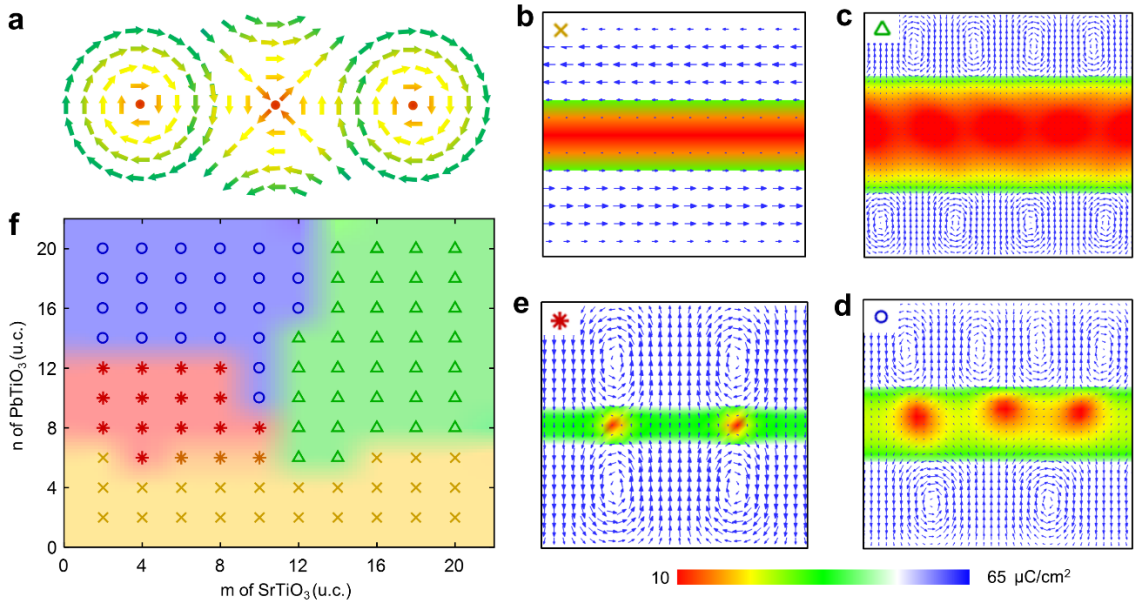

**Fig. 1 | Designing vortex-antivortex pair in (PTO)$_n$/(STO)$_m$ superlattice**. **a**, Schematic illustration of a topological antivortex sandwiched between two adjacent vortices. **b-e**, Four typical polar structures exist in (PTO)$_n$/(STO)$_m$ superlattices for different combinations of m and n, as predicted by phase field simulation: **b**, For 4-u.c. thick STO (m=4) sandwiched between two 4-u.c. thick PTO (n=4), anti-parallel *a*-domain is observed in PTO, while polarization in STO is negligibly small, exhibiting no nontrivial topological structure. **c**, For m=20 and n=10, vortex array emerges in PTO, while polarization in STO remains negligibly small. **d**, For m=10 and n=10, sign of topological structure appears in STO, with modestly increased polarization, while antivortex appears irregular. **e**, For m=4 and n=10, perfect antivortex array with relatively large polarization is observed in STO, sandwiched between two vortices in adjacent PTO. **f**, Phase field computed phase diagram of four typical polar structures in (PTO)$_n$/(STO)$_m$ superlattices, as represented by **b-e**.

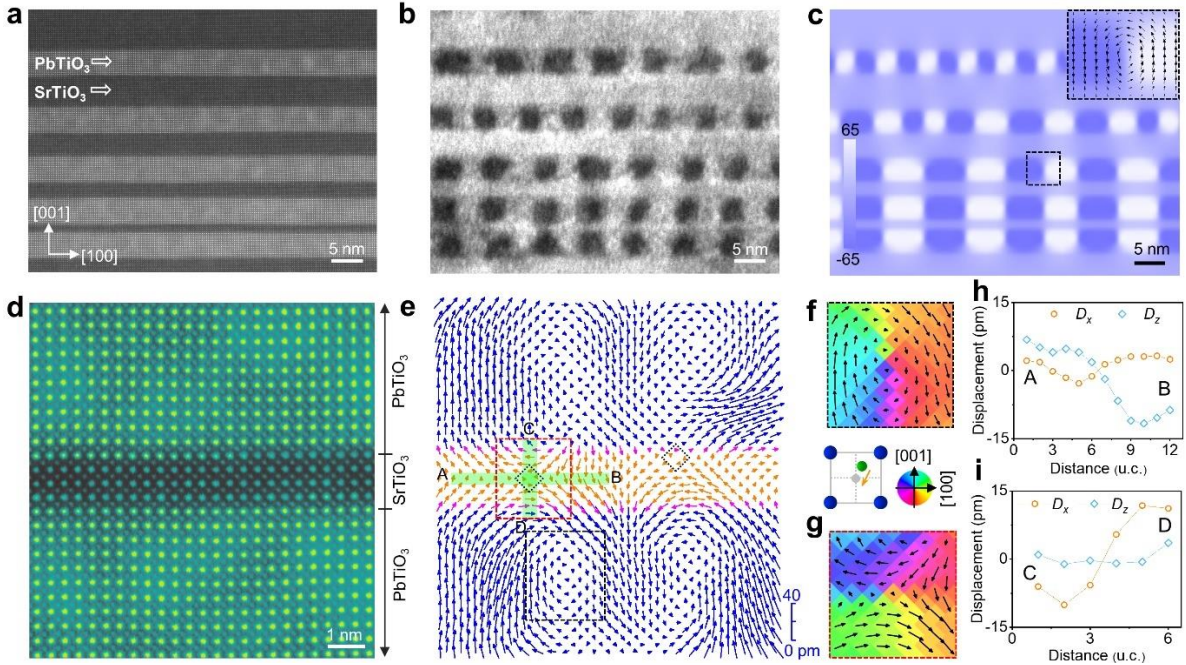

**Fig. 2 | Polar vortex-antivortex pairs in designed (PTO)ₙ/(STO)ₘ superlattice. a**, A low magnification HAADF image depicts STO layers with varying thicknesses (4, 7, 10 and 15-u.c.) sandwiched between 10-u.c. PTO layers. **b**, Dark field TEM image under two-beam conditions by selecting (002)$_{pc}$ **g**-vector (subscript pc denotes pseudo-cubic). The periodic array of bright and dark intensity modulation corresponds to vortex arrays within PTO layers. **c**, The spatial distribution of the out-of-plane polarization (unit: μC/cm²) calculated from phase field simulation. Inset: enlarged view of the polar vector configuration (black arrows). **d**, An atomically resolved HAADF image for a 4-u.c. thick STO sandwiched between adjacent 10-u.c. PTO layers. **e**, Map of polar vectors between cations extracted from the HAADF image depicts vortex-antivortex texture. The cores of antivortices are highlighted by the dotted diamond boxes. **f, g**, Enlarged views of polar vectors overlaid with polar angle variation taken from dashed highlighted rectangle boxes in **e** for vortex (black color) in PTO (**f**) and antivortex (red color) in STO (**g**), respectively. **h, i**, Variation of polar displacement components within antivortex structure along A-B (**h**) and C-D (**i**) directions, as marked in **e**. $D_x$ represents in-plane displacement and $D_z$ represents out-of-plane displacement.

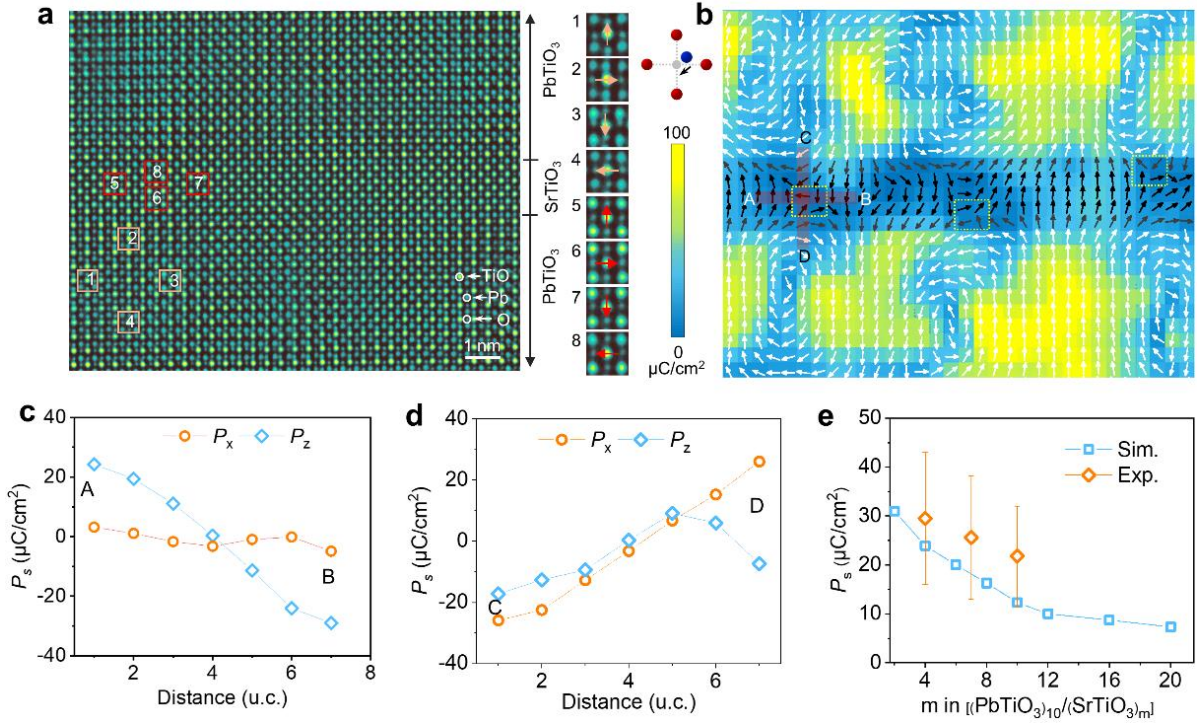

**Fig. 3 | Detailed polarization distribution of polar vortex-antivortex pairs in (PTO)₁₀/(STO)₄. a**, An atomically resolved iDPC image for 4-u.c. thick STO sandwiched between two 10-u.c. PTO layers, colored for clarity. From the enlarged views taken from the marked regions within PTO and STO, the atomic shift between cations and oxygen is visible with the naked eye. **b**, The corresponding unit-cell scale map of polarization vectors, calculated from the atomic displacements between cations and oxygen. Arrows denote the polarization orientation and the color represents the magnitude. The yellow dotted boxes highlight the locations of antivortex cores. **c, d**, Variation of polarization along A-B (**c**) and C-D (**d**) directions, as marked in **b**. **e**, The comparison of measured (orange) and phase field simulated (blue) average polarization versus m for (PTO)₁₀/(STO)ₘ. The error bar represents the standard deviation.

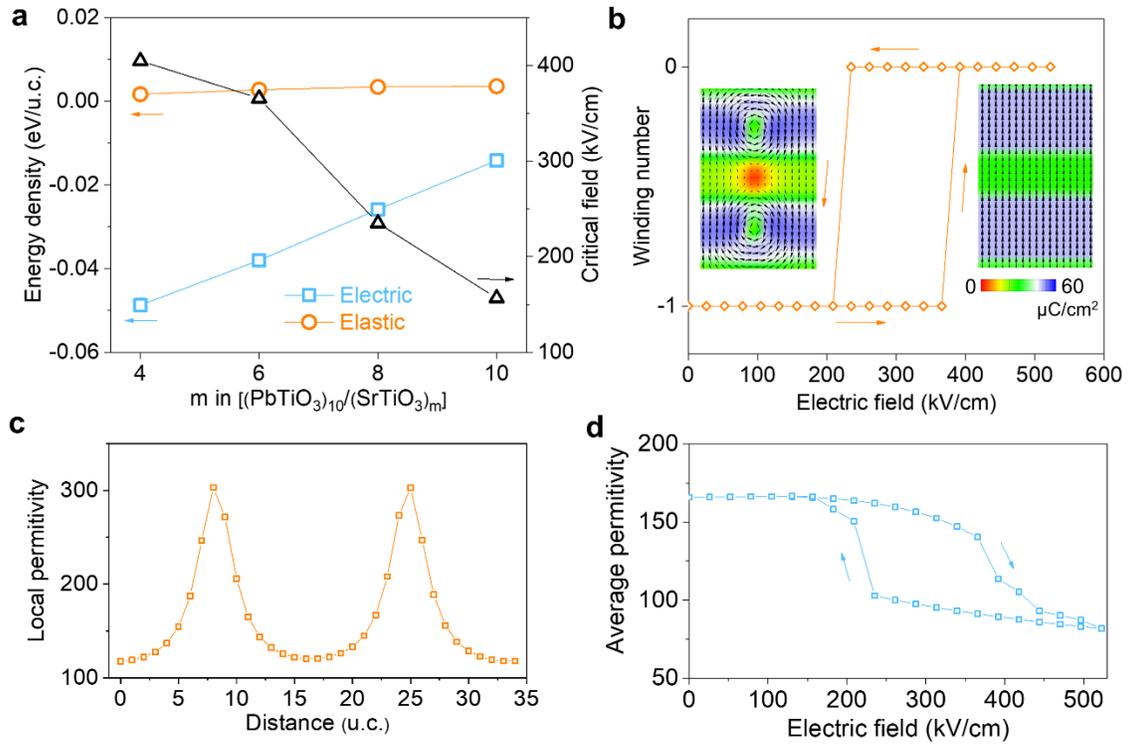

**Fig. 4 | Formation mechanism of antivortex in STO and its electric tuning via phase field simulation. a**, The electrostatic and elastic energy densities (on the left) for (PTO)$_{10}$/(STO)$_m$ heterostructures, and the critical electric fields (on the right) under which the antivortex disappears. **b**, The hysteresis loop of winding number of STO within (PTO)$_{10}$/(STO)$_6$ versus external electric field. The left and right insets show the polarization distributions of vortex-antivortex pair and single-domain state, respectively. **c**, The spatial distribution of local permittivity in the middle plane of STO across the antivortex cores. The two peaks indicate the significant increase of the permittivity at the antivortex cores. **d**, The average permittivity in the middle plane of STO versus the external electric field. The two abrupt changes of permittivity in the hysteresis loop are induced by the topological phase transition.

19